\documentclass[iop]{emulateapj}
\pdfoutput=1

\usepackage{amsmath}
\usepackage{aas_macros}

\newcommand\sB{{\textbf {\em B}}}

\newcommand\sV{{\textbf {\em V}}}

\newcommand\B{{\textbf {\em B}}}

\newcommand\divb{{\nabla \! \cdot \! \textbf{\em B}}}

\newcommand\jcoph{{J.~Comp.~Phys.}}

\begin{document}

\title{Phurbas: An Adaptive, Lagrangian, Meshless, Magnetohydrodynamics Code.  I. Algorithm}
\shorttitle{Phurbas MHD Code. I. Algorithm}

\author{Jason L.~Maron\altaffilmark{1}}
\email{{\tt jmaron@amnh.org}}
\author{Colin P.~M\textsuperscript{c}Nally\altaffilmark{2}}
\email{{\tt cmcnally@amnh.org}}
\author{Mordecai-Mark Mac Low\altaffilmark{2}}
\affil{Department of Astrophysics, American Museum of Natural History New York, NY, USA}
\email{{\tt mordecai@amnh.org}}

\altaffiltext{1}{current address: North Carolina Museum of Natural
  Sciences, Raleigh, NC, USA}
\altaffiltext{2}{Department of Astronomy, Columbia University, New
  York, NY, USA}
\begin{abstract} 

  We present an algorithm for simulating the equations of ideal
  magnetohydrodynamics and other systems of differential equations on
  an unstructured set of points represented by sample particles.
  Local, third-order, least-squares, polynomial interpolations
  (Moving Least Squares interpolations)  are calculated
  from the field values of neighboring particles to obtain 
       field values and
  spatial derivatives at the particle position.  Field
  values and particle positions are advanced in time with a second
  order predictor-corrector scheme. The particles move with the fluid,
  so the time step is not limited by the Eulerian
  Courant-Friedrichs-Lewy condition.  Full spatial adaptivity is
     implemented
  to ensure the particles fill the computational volume,
      which
  gives the algorithm substantial
  flexibility and power.  A target resolution is specified for each
  point in space, with particles being added and deleted as needed to
  meet this target.  Particle addition and deletion is based on a
  local void and clump detection algorithm. 
  Dynamic artificial viscosity fields provide stability to the integration.
  The resulting algorithm provides a robust solution for modeling flows
  that require Lagrangian or adaptive discretizations to resolve. This paper
      derives and 
  documents the Phurbas algorithm as implemented in Phurbas version 1.1.  A
  following paper presents the implementation and test problem results.
\end{abstract}


\section{Introduction}

\subsection{Context}
Our understanding of many astrophysical systems relies on the
simulation of magnetized plasmas.  As a result, much effort has been
made to develop tools to efficiently perform high-fidelity simulations of
them. Some of these tools have found broad application in other fields
of physics and engineering as well.

Many early methods for solving the equations of magnetohydrodynamics (MHD)
were based on fixed grids.  Discretizing the equations of
hydrodynamics or MHD on a fixed grid leads to an Eulerian method,
or a method written in terms of Eulerian derivatives.  Popular publicly
available codes with methods based on point values such as the Pencil
Code \footnote{See {\tt http://www.nordita.org/software/pencil-code/}}
\citep{2002CoPhC.147..471B}
and finite volumes, such as ZEUS \citep{2006ApJS..165..188H}, FLASH
\citep{2000ApJS..131..273F}, or Athena \citep{2008ApJS..178..137S} use
such methods.  Eulerian methods share the common property that the
discretized form of the governing equations is not Galilean
invariant. Though they still converge to the
correct solution, this does lead to two limitations at any finite
resolution.  First, the explicit integration time step constraint from
the Courant-Friedrichs-Lewy (CFL) condition depends on both the signal
speed and the flow velocity relative to the grid, not just the signal
speed.  Second, the numerical diffusion of the scheme, usually highly
nonlinear, also depends on the flow velocity relative to the grid.

A fixed grid approach thus has disadvantages
particularly where there are high-velocity bulk flows, collapsing flows,
or flows that generate localized fine structure.  For the latter
cases, adaptive mesh refinement \citep{1984JCoPh..53..484B} has been a
successful approach.  This method, while still Eulerian, uses refined meshes
to allow the spatial and temporal resolution to vary.  However, for problems with
significant bulk flows, it is of no help, as the same problems of
time step limitation and numerical diffusion apply as with uniform
grids.  A numerical viscosity dependent on the bulk flow
can be significant, because the growth of
instabilities from a marginally resolved mode in a method lacking
Galilean invariance will depend on the bulk velocity of the flow
across this grid.  The effects of this can be seen, for example, in
\citet{2008ApJ...675.1549C} and \citet{2009ApJ...697.1269J}.  To
circumvent the time step limit in disks treated with cylindrical or
spherical coordinates or in a shearing-sheet approximation where the
bulk flow is largely Keplerian and aligned with the grid, it is possible to add
a separate transport step to the method \citep{2000A&AS..141..165M}.
While this extra transport step improves the problems with numerical
diffusion, it does not fully cure the issue
\citep{2009ApJ...697.1269J,2010ApJS..189..142S}.

To escape these limits, it is necessary to move to a method formulated
in terms of Lagrangian (also known as covariant, comoving, convective,
advective, substantive, or material) 
 derivatives.\footnote{
Methods that solve Eulerian problems in a local
frame chosen to be comoving with the fluid in a locally average sense
also share in some of the advantages of this formulation.}
In contrast to Eulerian formulations, 
Lagrangian methods have three advantages.
Foremost, for problems with significant bulk flows, a purely Lagrangian formulation
has a significantly less stringent time step constraint from the signal
speed (the CFL condition).
This is because the time step in an Eulerian method depends on the maximum of the signal speed
and the flow speed, whereas in a purely Lagrangian method the time
step depends only on the local signal speed.
Relaxing this constraint becomes particularly important in the case of an
extended disk with supersonic differential rotation, where in an Eulerian
formulation the quickly orbiting inner regions constrain the time step
severely.
A second advantage of a Lagrangian method lies in the Galilean invariance
of the inevitable effects of numerical diffusion.
Though Galilean invariance itself can formally
be achieved in an Eulerian method
\citep{2010MNRAS.401..791S,2010MNRAS.401.2463R}, a Lagrangian formulation can
reduce the diffusivity further because it uses fewer time steps.
Finally, Lagrangian methods naturally focus resolution into regions of
fluid concentration, which are often, though not always, the regions
of greatest interest.  (We note that the adaptive, Lagrangian method
we describe here can also focus resolution to other, arbitrary regions
of interest.)

It is possible to write a comoving discretization in two ways.  First,
one can discretize the governing equations directly in terms of
Lagrangian time derivatives. Second, one can discretize in terms of
partial time derivatives around moving interfaces.  Historically, the most
popular approach has been the first, particularly when
used to build a meshless method. Recently, the second has been used,
with techniques based on a moving unstructured mesh with mesh
reconnection.

One of the earliest and most popular meshless schemes is Smoothed
Particle Hydrodynamics \citep[SPH;][]{1977AJ.....82.1013L,1977MNRAS.181..375G}.
SPH quickly gained popularity
as the advantages in numerical diffusion, local resolution scales, and local time step
advantages were realized \citep{1993A&A...268..391S}.  However, the basic SPH
algorithm has many shortcomings.  The foremost and most fundamental is the lack
of discrete, zeroth-order consistency in the SPH representation of a function. 
SPH interpolation fails
to reproduce even a constant function.
The importance of this consistency property in general meshless schemes has been 
pointed out by \citet{LiuJunZhang1995}.  This insight has been applied
to analysis of SPH by
\citet{Dilts1999,LiuLiuLam2003,Fries2004} and \citet{Quinlan2006},
among others. They find that the lack of zeroth-order consistency can
cause substantial gradient and value errors that do not converge with
increased particle number alone. 
The inability of SPH to effectively model subsonic turbulence has been blamed on
this lack of consistency by \citet{2011arXiv1109.4413B},
though the behavior in this regime depends strongly on, and can be significantly improved
by using a more modern formulation of the SPH artificial viscosity \citep{2011MNRAS.tmpL.377P}.

Resolution in SPH is further
limited by constant particle masses.  Some attempts at adaptive particle masses
have been made \citep{2002MNRAS.330..129K} but these suffer from difficulties
in specifying a well-posed scheme.  SPH in general handles
differing particle masses poorly, as the pairwise interparticle interactions allow heavy 
particles to penetrate though the fluid in a nonphysical manner. 
Similarly, the spatial
resolution in SPH is locally isotropic, even when the particle and mass
distribution is anisotropic. Attempting to relax this constraint leads to the
adaptive SPH scheme of \cite{1996ApJS..103..269S} and \citet{1998ApJS..116..155O}.

A grid that is both Lagrangian and has logically Cartesian structure is a simple choice, and
a logically Cartesian moving (Lagrangian) mesh has also been used to attempt to minimize
numerical diffusion \citep{1980ApJ...239..968N,1992ApJ...391..199F}.
\citet{1995ApJS...97..231G} and \citet{1998ApJS..115...19P} used a moving, logically Cartesian mesh
to provide adaptivity in collapsing flows.
However, this approach falls victim to several limits.
In many flows the cells eventually become long and thin, leading to large errors.
Also, the grid cannot follow 
rotation or turbulent
flows as it becomes tangled.

Unstructured, moving mesh methods with mesh reconnection have recently been
introduced in astrophysics.  The methods of
\citet{2010MNRAS.401..791S}, \citet{2011MNRAS.418.1392P}, 
\citet{2011ApJS..197...15D}, and \citet{2012arXiv1201.4873G} are
finite volume methods based on Voronoi 
tessellations. The mesh is defined by the Voronoi tessellation of a set of points
that move approximately with the mean motion of the fluid in the cell (though
formally any motion can be chosen).  
These methods can be described as Lagrangian though they 
calculate inter-cell fluxes with Eulerian Riemann problems stated in a locally comoving frame.
The connectivity of the mesh is dictated
by the Voronoi neighbor relation.  Fluxes between cells are calculated across
the moving cell faces.  
\citet{2010MNRAS.401..791S} and
\citet{2011MNRAS.418.1392P} describe a Galilean
invariant method.  The method of \citet{2011ApJS..197...15D} is not fully
Galilean invariant, but this is due to the formulation chosen for the slightly
more complicated relativistic hydrodynamic equations.
Both methods use an approximately comoving formulation in a significant sense.

This paper describes an adaptive, Lagrangian, meshless, collocation
scheme for MHD or similar sets of equations based on a point (not finite
volume or mass) discretization.  In what follows, we
refer to the discretization points as particles, following the
historical usage. However, these discretization
points do not in any sense represent identifiable masses or volumes
of the fluid.  They are simply moving points sampling continuous field
variables. 

In the next subsection 
we discuss prior work on related methods to
solve the MHD equations.
We then describe our algorithm, starting with an overview
(\S~\ref{sec_algorithm}) and then discussing specific numerical
aspects, such as the modeling of the function and the time update
(\S~\ref{sec_timederiv}), adaptive addition and deletion of particles
(\S~\ref{sec_regularizing}), 
explicit time step
limits (\S~\ref{sec_timestep}), 
and magnetic divergence
correction (\S~\ref{sec_magneticdivergence}). Finally we draw these
together with a summary of the algorithm (\S~\ref{sec_discussion}). 
In the next paper of this series \citep[][hereafter Paper II]{phurbatest} 
we present implementation details and present the
results of a suite of gas dynamical and MHD tests of the algorithm.

\subsection{Prior MHD Methods} \label{sec_priorwork}

Several attempts have been made to design an SPH-type scheme for MHD.  
The most successful and recent work by
\citet{2004MNRAS.348..123P,2004MNRAS.348..139P,2005MNRAS.364..384P}, and
\citet{2010MNRAS.401.1475P} resulted in an SPH MHD based on a form of the MHD
equations that is consistent with $\nabla \cdot \B \neq 0$ and a set of
artificial dissipation terms.  \citet{2007MNRAS.379..915R} developed a
variation based on representing the magnetic fields though Euler angles, which
allows a guaranteed $\nabla \cdot \B=0$ at the cost of disallowing tangled
field geometries \citep{2010MNRAS.401..347B}, severely limiting its applicability.
\cite{2009MNRAS.398.1678D} implement an SPH MHD in GADGET-3, without any
constraint on $\nabla \cdot \B$, but subtracting the numerical contribution of
$\nabla \cdot \B$ to the momentum equation.  We refer the reader to
\cite{2012JCoPh.231..759P} for a further overview of the attempts to design an
SPH MHD method.

Unfortunately all these SPH MHD methods suffer from the fundamental drawback of
SPH, that the SPH interpolant does not have a zeroth-order consistency property.  
This zeroth-order inconsistency means that for a disordered
set of SPH particles, a constant function cannot be reproduced by the SPH
representation of that function.  As the SPH representation of even a constant
function has significant positive and negative errors, it also has  significantly
non-zero derivatives. These errors make formulating an SPH MHD difficult.
Modifications of SPH to solve or work around the zeroth-order consistency
problem have been proposed. \citet{2001ApJ...561...82B} and \citet{2006ApJ...652.1306B} developed
an extension to SPH using a remapping strategy to increase the accuracy of SPH
estimates through regularizing the particle distribution, and applied it to MHD
shocks.
For hydrodynamics, \cite{1996PASA...13...97M} and \cite{2011MNRAS.413..271A} have proposed working around the
effects of the zeroth-order consistency problem for pressure forces only,
with an alternative derivation of the SPH pressure force. This comes at the price 
of sacrificing the local momentum conservation enjoyed by the classical formulation.
This also only treats the problem of spurious pressure forces
arising from the zeroth-order inconsistency, and does not lead to a
consistent interpolation 
of the pressure field or other fields.

It is also possible to construct a SPH MHD scheme using a Godunov
approach.
Godunov SPH was originally proposed for hydrodynamics, 
using Riemann problems to solve for the particle interactions.
Godunov SPH uses SPH interpolation for density 
(see Eq.~6 and Eq.~21 of \citeauthor{2002JCoPh.179..238I}
\citeyear{2002JCoPh.179..238I}, and 
Eq.~29 of \citeauthor{2011MNRAS.418.1668I} \citeyear{2011MNRAS.418.1668I})
\footnote{An earlier usage of Riemann solvers coupled with SPH is given by 
\citet{Parshikov:2000p7378}.}.
A Godunov SPH MHD implementation using Powell-type source terms and a tensile
correction was implemented by \citet{2011MNRAS.418.1668I}.
They point
out that all SPH-based MHD schemes that avoid tensile instability do not
exactly conserve momentum, energy, or both.
Similarly, \citet{2011MNRAS.414..129G} constructed an SPH-like scheme
(a weighted particle method)
with a consistent second order accurate formulation for derivatives, coupling
this with a pairwise Riemann-solver based interaction between particles to
yield an MHD scheme.  A Galilean invariant form of the \citet{2002JCoPh.175..645D}
hyperbolic-parabolic cleaning scheme was used to handle $\divb$ errors.  

However, these SPH-based methods again suffer from the zeroth order inconsistency of
the SPH interpolant, even though methods with a renormalized first derivative 
estimate have a consistent first derivative. This means that SPH interpolated 
fields (such as the density values) have significant
noise. To reduce the amplitude of the noise it is necessary to increase the number of
neighbors used in the kernel, which greatly increases the computational cost.  This means that
rigorous convergence studies, even in smooth flow, are not feasible with
methods based on SPH-type estimates.  In addition, SPH Riemann methods suffer a higher
computational cost in comparison to moving unstructured mesh Godunov
schemes, because of the requirement of a much higher number of Riemann problem solutions per
particle.

\citet{2011ApJS..197...15D} implemented a MHD scheme in their Voronoi
tessellation method, using a Dedner type hyperbolic divergence
cleaning method, but found it
difficult to manage $\divb$ errors when the mesh topology changes.
\citet{2011MNRAS.418.1392P} used a very similar approach, with apparently
much greater success in managing $\divb$ errors.
\citet{2012arXiv1201.4873G} add a source term to the induction equation to restore Galilean 
invariance if $\divb\ne 0$ and claim this greatly improves stability.

The method we describe here was inspired by the Gradient Particle Method of
\cite{2003ApJ...595..564M}, but removes the underlying instability 
present in that method (described in Appendix \ref{sec_stability}).  
A method particularly similar to
\cite{2003ApJ...595..564M}, but limited to hydrodynamics using a moving-least-squares fit was proposed by
\cite{Dilts1999,Dilts2000}.
Numerous related methods have been described in the literature on meshfree
or meshless methods.
The most closely related method is the Finite Pointset Method (FPM)
described by \cite{KuhnertThesis,FPM2002}, which is
not to be confused with either the similarly
named Finite Point Method of \citeauthor{otherFPM}
\citeyear{otherFPM}, 
or the equally similarly named Finite Particle Method of
\citeauthor{yetanotherFPM} \citeyear{yetanotherFPM}\footnote{The 
authors are of the opinion that enough numerical schemes have been
named FPM, and as the names are getting confusing the practice should cease.}.
FPM has limited adaptivity, is first order, and 
uses an upwinded formulation for hydrodynamics.
Similar to the method we describe, it is meshless, Lagrangian, has
particle addition and deletion, and uses moving least squares
interpolation. 

\section{Algorithm} \label{sec_algorithm}

    For specificity, we focus on using our method to solve
the equations of MHD.  These can be expressed using Lagrangian time
derivatives $D_t$, as
\begin{align}
D_t V_j &
= -\rho^{-1}\partial_j P \nonumber\\
& \qquad + \rho^{-1} \varepsilon_{jab}\varepsilon_{acd}(\partial_c B_d) B_b
+ G_j,
   \label{eqmomentum}
\\
D_t B_j &
= B_i \partial_i V_j - B_j \partial_i V_i, \label{eqinduction}
\\
D_t \sigma &= -(\sigma +P )\partial_i V_i  \label{eqenergy} \\
D_t\rho &= -\rho\partial_i V_i, \label{eqmass}
\end{align}
where $V$ is the velocity, $B$ is the magnetic field, $\sigma$ is the
internal energy volume density,
$P$ is the pressure,
$\rho$ is the density, $G_j$ is a vector component of a body force,
and the Einstein summation rule is assumed.
We note that Phurbas is relatively insensitive to the exact form of the equations solved and 
variables chosen. For example, energy variables other than the 
internal energy per volume could be used.
In Appendix~\ref{app_secondderiv} we give the second time derivatives of these equations for use
in the time update.
These equations require the addition of an equation of state, such as a gamma-law 
$P=(\gamma-1)\sigma$, though the equation of state is arbitrary.

The MHD equations~(\ref{eqmomentum})--(\ref{eqmass}) are
solved on an adaptive set of particles, each particle carrying values for the
field variables $ \rho, \sV, \sB, $ and $\sigma$.  Particles move in
the frame of the fluid with the local fluid velocity $\sV.$ Field
variables evolve in the frame of the particle, so the evolution
equations are most naturally expressed using the
Lagrangian form for the time derivatives in the MHD equations.

The equations of MHD as stated are ill suited to the numerical scheme
we will use.  For the discretization used in Phurbas, we require a
system of equations in which short wavelength perturbations decay.
Appendix~\ref{sec_stability} demonstrates this for a model
advection-diffusion equation.  To ensure decay of such perturbations,
we introduce artificial dissipation terms to the analytic form of the
equations before discretizing. 
These modifications are in the form of a bulk
viscosity, and mass and thermal diffusions.  Formally, it is this modified version of
the MHD equations from which Phurbas computes approximate numerical
solutions.  The MHD equations, reiterated with the addition of the
stabilizing terms, and associated fields are:
\begin{align}
D_t V_j &
= -\rho^{-1}\partial_j P + \rho^{-1} \varepsilon_{jab}\varepsilon_{acd}(\partial_c B_d) B_b \nonumber\\
& \qquad
+ G_j + \partial_j\left( (\zeta_s + \zeta_l)\partial_i V_i\right),
   \label{eqmomentumstab}
\\
D_t B_j &
= B_i \partial_i V_j - B_j \partial_i V_i + \xi_j, \label{eqinductionstab}
\\
D_t \sigma &= -(\sigma +P )\partial_i V_i  + (\zeta_s + \zeta_l)(\partial_i V_i)^2 \nonumber\\
& \qquad
 + H_\sigma \rho \partial_i(\zeta_s \partial_i \frac{\sigma}{\rho}),  \label{eqenergystab}\\
D_t\rho &= -\rho\partial_i V_i + H_\rho \partial_i( \zeta_s \partial_i \rho) , \label{eqmassstab}\\
D_t \zeta_l &=  \partial_i(\kappa_\zeta \partial_i \zeta_l) +\frac{1}{\tau_l}\lambda c_\mathrm{max} - \frac{1}{\tau_l}\zeta_l, \label{eqzetal}\\
D_t \zeta_s &=  \partial_i(\kappa_\zeta \partial_i \zeta_s) +\frac{1}{\tau_{s+}}S_s - \frac{1}{\tau_{s-}} \zeta_s, \label{eqzetas}
\end{align}
where $\lambda$ is the Nyquist length, and
$c_\mathrm{max} = \sqrt{c_s^2 +v_a^2}$ is the maximum signal speed, where $c_s$ is the sound speed and $v_a$ is the Alfv\'{e}n speed.
Bulk viscosity fields $\zeta_l$ and $\zeta_s$ are introduced to handle
the general flow ($\zeta_l$), and shocks and other 
discontinuities ($\zeta_s$).
The action and parameterization of these fields are described in \S~\ref{sec_bulkviscosities}.
$H_\sigma$ and $H_\rho$ are constants that specify the strength of
mass and thermal conductivities in continuity and energy equations, while 
$\xi_j$ is the term representing diffusion of magnetic divergence defined in \S~\ref{sec_magneticdivergence}.

To evolve the field variables in time, we evaluate
Equations~(\ref{eqmomentum})--(\ref{eqmass}) for the time derivatives,
requiring values for the field variables and their spatial derivatives
at the position of each target particle. We obtain this information by
fitting a third-order, three-dimensional (3D) polynomial 
   to the set of
values carried by
the neighboring particles, using the procedure described in
\S~\ref{sec_timederiv}.  The resulting polynomial coefficients allow
us to compute the field
value and its first, second, and third derivatives at the position of the particle, enabling
evaluation of the Lagrangian time derivatives.  Those in turn are used
to update
the field variables with a predictor-corrector time step scheme
described in \S~\ref{sec_timeupdate}.

A particle-based algorithm such as this one has a dynamically evolving
spatial resolution.  It turns out to be central to the stability and
accuracy of the method that the particle distribution not have voids
within which the fields cannot be accurately fit. We create and delete
particles as necessary to eliminate such voids, while avoiding
particle clumps.  This further allows us to adaptively satisfy any
user-specified physical resolution requirement, as well as to
eliminate unnecessary particles (\S~\ref{sec_regularizing}). We force
the resolution to always exceed a spatially and temporally variable
target resolution $\lambda({\textbf{\em x}},t)$.  Effectively, the particles can
represent the field variables in the same manner as a grid with
effective resolution $\lambda$ at each point.  The resolution
requirement can be specified depending on the physics requirements of
the problem at hand, so long as it remains reasonably smooth.

The Phurbas discretization is based on point values, not
finite volumes or finite masses.
As such, the discretization used to calculate spatial derivatives and
advance the solution in 
time does not define a value for volume-integrated
quantities, including volume-integrated, conserved quantities. 
To define these quantities, another discretization would need to be added to 
obtain a multidimensional quadrature from the unstructured set of samples.
For example, Voronoi cell volumes could be used to calculate a nearest-grid-point 
interpolation for a Riemann-sum approximation to a volume integral.
Alternatively, using a point density approximated from the number of neighboring points 
in some small radius as a weighting, a Monte-Carlo type volume
integral approximation could be used.

As the magnetic field evolves, discretization error generates spurious
magnetic divergence $\divb$. 
By dropping the physically vanishing term $-\divb$ when deriving the Lagrangian induction equation
from the usual Eulerian form expressed
with a partial time derivative, we have made any $\divb$
present in the field into a passively advected scalar.
A consequence of this choice of the canonical Lagrangian form is that
our MHD equations, by omitting a term that is physically zero, are
precisely the same as a form that is claimed in other works to include
an extra source term: the 
same result has been proposed with a source
term by \cite{2000JCoPh.160..649J}, and derived from the relativistic
form of energy-momentum conservation and relativistic electromagnetic
theory by \cite{2001JCoPh.172..392D}. In the latter paper, it is shown to be the
Galilean invariant momentum and energy conserving form for the MHD
equations in the case when $\divb$ is present.  We note that $\divb$
of nonphysical origin is easier to numerically handle in the Lagrangian 
form of the MHD equations, as the presence of $\divb$ errors does not feed back
into violations of energy and momentum conservation by itself.  Unlike
in SPH MHD \cite{2012JCoPh.231..759P}, we do not require source terms
in the momentum and energy equations, as our discretization does not
suffer from the tensile instability as in SPH.
The diffusive correction described in \S~\ref{sec_magneticdivergence} adds a term to
the magnetic equation that diffuses $\divb$.

\section{Time Evolution} \label{sec_timederiv}

We evolve the field variables forward in time by evaluating the MHD
equations~(\ref{eqmomentum})--(\ref{eqmass}) at the position of each
particle.  We do this by constructing a local
approximation of the field variables at that position, derived from a
spatial fit to the values of the field variables on neighboring
particles~(\S~\ref{sec_matrix}). This allows us to compute the values and
spatial derivatives of the field variables at the position of the
particle.  We choose for the form of the continuous approximation a
3D, third-order, polynomial. We further develop a system of particle
weights that enhances the accuracy of the fit (\S~\ref{sec_weights}).
Once we have evaluated the Lagrangian time derivatives from the MHD
equations, the field variables and particle positions are updated in
time with a predictor-corrector method (\S~\ref{sec_timeupdate}).

\subsection{Moving Least Squares Procedure} \label{sec_matrix}

Phurbas uses a moving least squares fitting procedure in two versions.
First, to approximate derivatives of the dynamical field variables on the right
hand sides of the governing equations, 
moving least squares {\em interpolants} are used.
These are polynomial approximations that are forced though (interpolate) the central particle value.
Second, Phurbas uses moving least squares {\em fits} to initialize newly created particles. 
These are polynomial approximations that are not forced though any particle values,
and hence provide a smooth approximation to the field
values where no particle currently exists.

In the language of \citet{LancasterSalkauskas1981} the first version
is an Interpolating Moving Least Squares procedure,
and the second is a Moving Least Squares procedure. 
In addition to \citet{LancasterSalkauskas1981}, discussion at length
can be found, for example, in
\citet{Belytschko1996}, \citet{Dilts1999}, or \citet{Fries2004}.

For the purposes of Phurbas, we can follow the description by \citet{Liszka1996263}, which leads to 
what has become known as a Generalized Finite Difference Method \citep{Liszka198083}.
Phurbas uses both the Nodal Approximation described in \citet[section 2.1.1]{Liszka1996263} and the
Pointwise Approximation in \citet[section 2.1.2]{Liszka1996263}.
We briefly expand on those descriptions here for clarity.

In one dimension, we start with a function that we wish to discretize,
defined at some set of points $g = g(\mathbf{x})$. 
To approximate the function at a point $\mathbf{x}_0$, we select a set of nearby points
$\{\mathbf{x}_i\}$, and then write the series approximation using $n$
polynomial terms $p_i(\mathbf{x})$,
\begin{align}
q(\mathbf{x}) = \sum_{i=0}^{n} a_i p_i(\mathbf{x}- \mathbf{x}_0)
\label{genapprox}
\end{align}
so that $q(\mathbf{x}) \approx g(\mathbf{x})$.
If the functions $p_i$ are selected as polynomials,
\begin{align}
p(\mathbf{x}) = \left[1,x,y,z,x^2,y^2,z^2,xy,xz,x^3,...\right]
\end{align}
then this approximation is a Taylor series about $\mathbf{x}_0$, as in
Equation~(\ref{genapprox}) we have shifted the  
polynomials $p$ to be centered on $\mathbf{x}_0$.
The coefficients $a_i$ are then the value and derivatives (multiplied by a Taylor series coefficient) of
the approximation $q(x)$.

If the function $g$ is defined at $\mathbf{x}_0$ we can 
reduce the approximation to a special case, called the Nodal
Approximation by \citet[section 2.1.1]{Liszka1996263}.  If
we fix the coefficient $a_0 = g(\mathbf{x}_0) = q(\mathbf{x}_0)$, then
only the coefficients $a_i$ for $i > 1$  need to be determined, and the approximation becomes
\begin{align}
q(\mathbf{x}) = g(\mathbf{x}_0) + \sum_{i=1}^{n} a_i p_i(\mathbf{x}- \mathbf{x}_0)
\label{nodalapprox}
\end{align}
We choose to use a number of neighboring points $n$ greater than the number of undetermined polynomial coefficients $m$.
As Equation~(\ref{nodalapprox}) is then overdetermined, we seek a solution in the least-squares sense.
That is, the solution for $a_i$ should minimize the quadratic form
\begin{align}
J=\sum_{j=1}^{n}W(\mathbf{x}_j-\mathbf{x}_0)\left(g(\mathbf{x}_j) - q(\mathbf{x}_j)\right)^2,
\label{quadform}
\end{align}
where $W$ is a weight function described below.
We can rewrite this set of equations in matrix form by defining
\begin{align}
\mathbf{g}^\mathbf{T} &= \left[g(\mathbf{x}_1)- g(\mathbf{x}_0),g(\mathbf{x}_2)- g(\mathbf{x}_0),g(\mathbf{x}_3)- g(\mathbf{x}_0),...\right],\\
\mathbf{P} &= \left[
\begin{array}{cccc}
p_1(\mathbf{x}_1 - \mathbf{x}_0) & p_2(\mathbf{x}_1 - \mathbf{x}_0) & \dots & p_m(\mathbf{x}_1 - \mathbf{x}_0)\\
p_1(\mathbf{x}_2 - \mathbf{x}_0) & p_2(\mathbf{x}_2 - \mathbf{x}_0) & \dots & p_m(\mathbf{x}_2 - \mathbf{x}_0)\\
\vdots & \vdots & \ddots & \vdots \\
p_1(\mathbf{x}_n - \mathbf{x}_0) & p_2(\mathbf{x}_n - \mathbf{x}_0) & \dots & p_m(\mathbf{x}_n - \mathbf{x}_0)
\end{array}\right],
\end{align}
and
\begin{align}
\mathbf{W} = \left[
\begin{array}{cccc}
W(\mathbf{x}_1-\mathbf{x}_0) & 0 & \dots & 0\\
          0         &W(\mathbf{x}_2-\mathbf{x}_0) &  \dots & 0\\
\vdots & \vdots & \ddots & \vdots \\
0    &     0 &   \dots         &W(\mathbf{x}_n-\mathbf{x}_0)\\
\end{array}\right].
\end{align}
Then Equation (\ref{quadform}) can be written
\begin{align}
J = (\mathbf{P}\mathbf{a} - \mathbf{g})^\mathbf{T} \mathbf{W} (\mathbf{P}\mathbf{a} - \mathbf{g}).
\end{align}
If we define
\begin{align}
\mathbf{A} = \mathbf{P}^\mathrm{T} \mathbf{W}\mathbf{P}\\
\mathbf{B} = \mathbf{P}^\mathrm{T} \mathbf{W}
\end{align}
then minimizing $J$ as in \cite{Belytschko1996} we obtain
\begin{align}
\mathbf{a} =\mathbf{A}^{-1}\mathbf{B}\mathbf{g}
\label{eq:matrix}
\end{align}
and
\begin{align}
\mathbf{a}^\mathbf{T} = \left[\frac{\partial q}{\partial x},\frac{\partial q}{\partial y},\frac{\partial q}{\partial z}, 2\frac{\partial^2 q}{\partial x^2},...\right]  
\end{align}
gives the derivatives of the interpolating moving least squares approximation.
 
The second form, the Pointwise Approximation, is defined everywhere,
not just where there is a particle. 
This form is used when adding new particles. 
The coefficient $a_0$ is left free, the approximation is
Equation~(\ref{genapprox}), and now 
at an arbitrary point $\mathbf{x}$ the approximation yields:
\begin{align}
\mathbf{g}^\mathbf{T} &= \left[g(\mathbf{x}_0),g(\mathbf{x}_1),g(\mathbf{x}_2),g(\mathbf{x}_3),...\right],\\
\mathbf{P} &= \left[
\begin{array}{cccc}
p_0(\mathbf{x}_0 - \mathbf{x}) & p_1(\mathbf{x}_0 - \mathbf{x}) & \dots & p_m(\mathbf{x}_0 - \mathbf{x})\\
p_0(\mathbf{x}_1 - \mathbf{x}) & p_1(\mathbf{x}_1 - \mathbf{x}) & \dots & p_m(\mathbf{x}_1 - \mathbf{x})\\
\vdots & \vdots & \ddots & \vdots \\
p_0(\mathbf{x}_n - \mathbf{x}) & p_1(\mathbf{x}_n - \mathbf{x}) & \dots & p_m(\mathbf{x}_n - \mathbf{x})
\end{array}\right],
\end{align}
and
\begin{align}
\mathbf{W} = \left[
\begin{array}{cccc}
W(\mathbf{x}_0-\mathbf{x}) & 0 & \dots & 0\\
          0         &W(\mathbf{x}_1-\mathbf{x}) &  \dots & 0\\
\vdots & \vdots & \ddots & \vdots \\
0    &     0 &   \dots         &W(\mathbf{x}_n-\mathbf{x})\\
\end{array}\right].
\end{align}
The vector of coefficients then yields
\begin{align}
\mathbf{a}^\mathbf{T} = \left[q(\mathbf{x}),\frac{\partial q}{\partial x},\frac{\partial q}{\partial y},\frac{\partial q}{\partial z}, 2\frac{\partial^2 q}{\partial x^2},...\right]  .
\end{align}

The coefficients vector has either 19 or 20 coefficients, depending on
the approximation, so the
solution requires inversion of either a $19 \times 19$ or $20 \times 20$ matrix
 ${\textbf{A}}$.  We use an LU decomposition and back substitution procedure
\citep[e.g.][p. 32]{1992nrfa.book.....P} to solve
Equation~(\ref{eq:matrix}). The derived polynomial coefficients yield
the values of the field variables and their derivatives of first,
second and third order, from which we construct the first- and
second-order time derivatives of the field variables.

We have experimentally found that the number of particles included 
in the evaluation sums for the matrix
coefficients should comfortably exceed the number of terms in the
polynomial. The choice of how many particles to include is based on a compromise
between lack of statistical significance and computational
impracticality.  The radius $r_f$ of the sphere encompassing the
particles included should also be large enough to justify a third-degree
interpolation, about twice the characteristic inter-particle separation
$\lambda$.  For a uniform particle density of one particle within each
volume $\lambda^3$, 
a sphere with $r_f = 2
\lambda$ encloses $\sim 34$ particles. However, this leaves little room to
account for non-uniform particle densities. A sphere with $r_f = 3
\lambda$ encloses $\sim 110$ particles, which weighs on the cost of
calculating the matrix coefficients. A radius as large as this also
invites higher-order structure to erode the interpolation.  In the end we choose
to use 
\begin{equation} 
r_f = 2.3 \lambda,
\end{equation} 
corresponding to $\sim 51$ particles.
We have not yet derived a rigorous lower bound to the required number of
neighbors to use.
For computational cost, we find that a third-order fit has
computational cost comparable to the
other operations that occur in the time step, while a fourth-order
fit is substantially more expensive.
We term the sphere of radius $r_f$ around a fit center the neighbor
sphere. 

The target resolution
$\lambda ({\textbf{\em x}},t)$ is a property of the location of the fit center, which
may or may not be  
centered at the location of a particle.
In practice, as the interpolation is centered at the location of an existing particle, 
then $\lambda$ for the interpolation is taken as the $\lambda$ of that particle.
If a fit is being used to generate new values for the addition 
of a particle, the $\lambda$ of the particle
that triggered the creation of the new particle is used.

\subsection{Weights} \label{sec_weights}
In the moving least-squares procedure, each neighbor particle $j$ has a weight $W_j$.
There is significant freedom to choose the form of the weights $W_j$,
and no
rigorous theoretical framework exists under which an optimal choice
can be made.  As a first approximation, we choose weights
that emphasize particles close to the center of the
neighbor sphere, so that the local least squares approximation varies more smoothly as the
target position is changed.  The weighting function is a piecewise linear function of the distance
of particle $j$ from the fit center $r_j$, given as:
\begin{equation}
W_{j} = \begin{cases}
1 & \text{if }0\leq r_j< r_w \\
1-\frac{3}{5} \left(\frac{r-r_w}{r_f-r_w} \right) & \text{if } r_w \leq r_j \leq r_f,
\end{cases}
\end{equation}
where $r_w = \frac{3}{2}\lambda$. 
$W_{j}$ has a nonzero value at the edge of the neighbor sphere so as to not exclude any particles from the fit.

\subsection{Time Update} \label{sec_timeupdate}

The field variables are evolved in time with a Hermite predictor-corrector
scheme based on the first- and second-order
Lagrangian time derivatives.
A derivation of the scheme is presented in Appendix
\ref{app_timeintegration}, as it has not previously been described in
the literature. 
The interpolation procedure in \S~\ref{sec_matrix}  
for time step $i+1$ is done on the predicted 
values $q_{p,i}$ computed in time step $i$, 
yielding the time derivative values $D_t q_{p,i}$ and $D_{tt}
q_{p,i}$ needed for the correction in time step $i$,
as well as for the prediction of
time step $i+1$. 

We begin by
extrapolating forward from time $t_i$ to time $t_{i+1}$, 
over the time interval $\Delta t = t_{i+1} - t_i$, to make a prediction
\begin{equation} \label{predict}
q_{p,i+1} = q_{c,i} + D_{t}q_{p,i} \Delta t + \frac{1}{2}D_{tt}q_{p,i} \Delta t^2,
\end{equation}
based on a Taylor series expansion around $q_{p,i}$, 
using the corrected value from the previous time step $q_{c,i}$.
We then evaluate the time derivatives by interpolation on the
predicted fields $q_{p,i+1}$ at time $t_{i+1}$, and correct the prediction to derive
the corrected value at $t_{i+1}$,
\begin{align}
\label{correctorstepeq}
q_{c,i+1} =& q_{c,i} + \frac{1}{2}(D_t q_{p,i} + D_t q_{p,i+1} ) \Delta t \\
& +\frac{1}{12}(D_{tt} q_{p,i} - D_{tt} q_{p,i+1}) \Delta t^2 .\nonumber
\end{align}

The particle positions ${\textbf {\em x}}$ are evolved using third-order time
information
$D_{ttt} {\textbf{\em x}} = D_{tt} \sV$
as well.
This allows us to use a third-order predictor 
of the form
\begin{align}
x_{p,i+1} = &x_{c,i} + V_{c,i} \Delta t \\
&+ \frac{1}{2}D_{t}V_{p,i} \Delta t^2 + \frac{1}{6}D_{tt}V_{p,i} \Delta t^3 \nonumber
\end{align}
and to correct it to the final value
\begin{align}
x_{c,i+1} = & x_{c,i} + \frac{1}{2}(V_{c,i} + V_{c,i+1} ) \Delta t \nonumber \\
& +\frac{1}{10}(D_{t} V_{p,i} - D_{t} V_{p,i+1}) \Delta t^2 \\
& +\frac{1}{120}(D_{tt} V_{p,i} + D_{tt} V_{p,i+1}) \Delta t^3 \nonumber
\end{align}

\section{Regularizing the Particle Distribution} \label{sec_regularizing}

Our algorithm relies on discretization over Lagrangian sample points.  These
points are not arrayed on a grid, nor are they connected by mesh edges
as in the AREPO code \citep{2010MNRAS.401..791S}, so this is a meshless method.
During evolution, we require 
that the particles should maintain a
distribution such that there are no voids larger than the target
local resolution $\lambda({\textbf{\em x}},t)$ and no excessive point concentrations
within the scale $\lambda$. 
The requirement of no voids is 
introduced to ensure that every fit sphere of radius $r_f$ has enough points to 
perform the moving least-squares procedure and that the fit spheres overlap sufficiently
so that the set of fit spheres covers the entire simulation volume.
The requirement of no point concentrations requires the removal of 
excess points. 
We implement these requirements by adding and deleting
particles as needed. 
Satisfying these requirements confers the great benefit of making
the code fully adaptive, since the user can dynamically choose the function
$\lambda({\textbf{\em x}},t)$ as required by the physics of the
problem, so long as it is reasonably smooth in space and time.

The addition and deletion algorithm begins with the assembly of all neighbors $i$ within the
neighbor sphere of radius
$r_f$ around a particle $j$, along with their associated target
resolutions $\lambda_i.$ Voids within the neighbor sphere are
identified using the method described in \S~\ref{sieve}. Any voids
identified are reported as candidates for particle creation. 
Conversely, if a particle $j$ has a mutual nearest neighbor that is too
close (see \S~\ref{sec_clumpbusting}), one of the two particles is deleted. 
Duplicate voids and clumps are pruned from the
global list prior to the particle creation and deletion described in
\S~\ref{creation}. 

\subsection{Voids} \label{sieve}

To check for a void at a point in space with position ${\textbf {\em x}},$ we identify the nearest particle $i$,
which is located at position ${\textbf {\em x}}_i$ and has a resolution scale $\lambda_i.$
The distance between ${\textbf {\em x}}$ and the particle normalized by the resolution
scale is then
\begin{equation} \label{eq:xvoid}
x_{{ void}} = \frac{|{\textbf {\em x}} - {\textbf {\em x}}_i|}{\lambda_i}
\end{equation}
As a resolution condition, we then choose the condition that if
\begin{equation}
x_{{ void}} > c_{{ void}}, \label{eq:void}
\end{equation}
for a constant $c_{{ void}}$,
the space around ${\textbf {\em x}}$ is indeed too sparsely populated, indicating a need for
particle addition.  

To heuristically derive $c_{ void}$, we
consider an arrangement of particles on the hexagonal lattice
representing the tightest possible packing of spheres centered on the
particles.  If the particle density is one particle per volume
$\lambda^3$, then the spheres' centers will be 
separated by a distance $d_p =  2^{1/6} $.  This represents
the most efficient possible
  filling of the region with particles.  
Since any real, fluctuating, particle distribution will require more
particles to fully resolve the field, we set
\begin{equation}
c_{ void} = 0.73 d_p 
\end{equation}
so that with a disordered particle set we sample more density than 
would be required with the ideal ordered particle set.

To identify unique voids, we first identify the most egregious void
within the neighbor sphere of each particle, and then check to see if
that void violates the condition given by Equation~(\ref{eq:void}).  If
it does we add a particle as described below in Section~\ref{creation}.
We begin by examining the space in the vicinity of existing
particles. We construct a 3D cubic grid 
with side length $2r_f$ containing $9\times 9\times 9$
grid points, centered on the target particle position ${\textbf{\em x}}_j$.
This grid covers the volume of the neighbor sphere. For each
grid point, the normalized distance $x_{ void}$ to all neighboring
particles can be calculated using Equation~(\ref{eq:xvoid})
and the minimum value chosen.  If the maximum value on the grid of
$x_{ void} > c_{void}$, the position of the grid
point with the maximum value is reported as a
 candidate void for particle creation.

To speed up the calculation, we sieve the grid points lying within the
neighbor sphere.  We begin the search by
initializing a large value on each grid point for the minimum value of
$x_{ void}$ for that grid point.  
We then proceed by selecting each particle $i$ in turn, and looping
over all grid points.
For each grid point, we calculate the normalized distance $x_{ void}$
to the particle $i$.
If its value for the grid point is less than the current minimum value on that point, we
replace it with the newly calculated value for particle $i$. If the
new value is less than $c_{void}$, that grid point can be eliminated
from the active list of candidates for void identification.  We then
move to the next particle and calculate its distance to the remaining
active grid points, repeating the above procedure.  After all
particles have been sieved, if any grid points remain as void
candidates, we report the one with the maximum $x_{ void}$ as a candidate for void creation.
To hasten the operation, we first sieve the particles within 
$\lambda$
from the target position, then those within $(3/2) \lambda$, and then the
remaining particles, where $\lambda$ is the value for the target position.

\subsection{Clumps} \label{sec_clumpbusting}

As the particles move, random fluctuations will move them closer or
farther from their neighbors. If two particles approach each other too
closely compared to $\lambda$, they are essentially sampling the same
field variable information, and so are redundant.  Because there are
no restoring forces in the algorithm to separate nearby particles, we instead remove
any particle clumps of this sort, saving the computational cost of
evolving the redundant particles. 
The question then remains of how to determine when a clump has formed.  

To do this, we define a scaled distance between two particles:
\begin{equation}
r_{ij}^2 = \frac{({\textbf {\em x}}_i - {\textbf {\em x}}_j)^2}{\lambda_i \lambda_j}.
\end{equation}
The nearest neighbor $i$ to the target particle $j$ is determined. In
turn its nearest neighbor is found. 
If they are mutual nearest neighbors and if 
\begin{equation}
r_{ij} < c_{ clump}
\end{equation}
they are candidates for deletion.
We find that a value of
\begin{equation}
c_{ clump} = 0.12 d_p 
\end{equation}
is suitable to prevent over-resolution.  Among those two particles, we
delete the one which was more
recently created, retaining the particles with longer history to minimize the
numerical diffusion from adaption.  

If particles are to be deleted, this is done so without considering whether
that particle triggered the proposed addition of a particle (that is,
whether it is in a clump on the edge of a void). However, any proposed
addition resulting from the processing of that particle is still considered.

\subsection{Particle Creation and Deletion} \label{creation} 

The first examination of all the active particles results in a
proposed list of positions requiring particle addition, accompanied by
the radius of the void detected. These
proposals overlap, as each void may be detected by more than one
particle. The list is exchanged by processes handling neighboring
spatial domains, so that each process has a list of all the proposed
additions within a distance $r_f$ of its boundary.  The proposed
addition list is then pruned, to select one position in which to add a
particle within each void radius. 
To do this, each particle addition proposal is
compared to all other proposals within that spatial domain.  If any
other proposed location lies within its void radius, the values of the
void radii are compared, and the proposal with the smaller void
radius is rejected.

We then create particles at the successfully proposed positions. 
Particles in this algorithm represent sample points, not discrete
parcels of gas.  When we add particles, we are just sampling the
continuous field variables at new positions.  Therefore,
considerations of conservation do not enter this process, unlike in
particle-splitting methods used in SPH
\citep[e.g.][]{2002MNRAS.330..129K}. 

The task of creating new particles needs to be load balanced among
processors in order to 
handle situations where the memory required for new particles represents a
large fraction of the total free memory in the particle arrays. 
The new
particles are then initialized in free spaces, on the processors to
which they have been assigned by the addition load balance procedure.
As we have now deleted some particles, and added others to essentially
random processors, a new load balance may be calculated among all particles,
and the neighbor search data structure must be updated.
Doing this on the entire particle list brings
the new particles to optimal positions on the processors and
provides neighbor information for the subsequent processing stages.

New particles are initialized using a third order moving least squares fit, as opposed
to an interpolation (\S \ref{sec_matrix}). This fit is centered on the position of the new particle.

\section{Time Steps} \label{sec_timestep}

The time step for each particle is set by taking the minimum of five
criteria.
These are evaluated at the phase where new time derivatives are computed.
The basic limit is the CFL condition for the
stability of a forward-time-centered-space 
discretization. It is used here without explicit derivation as the
general principle applies that the maximum stable
time step must be short enough that a signal cannot cross a distance
exceeding the local resolution $\lambda$,
\begin{equation}
\Delta t_\mathrm{CFL} = C_\mathrm{CFL}\frac{\lambda}{\sqrt{c_s^2+v_A^2}}
\end{equation}
where $\Delta t_\mathrm{CFL}$ is the CFL time step, $C_\mathrm{CFL}$
is the Courant number, which we usually take to be $0.3$, $c_s$ is the
sound speed, and $v_A$ is the Alfv\'{e}n speed. Note that, unlike an
Eulerian code, the flow velocity does not enter this equation.

The use of the bulk viscosity fields $\zeta_s$ and $\zeta_l$ require
an appropriate time step constraint related to the diffusion term they
introduce in the momentum equation,
\begin{equation}
\Delta t_{\zeta} = \frac{C_\mathrm{CFL}}{\pi^2 (\zeta_s+\zeta_l)}.
\end{equation}
Another time step constraint
of the same form
is applied based on the need for a sufficient number of time steps
during a compression or expansion to allow for particle addition and
deletion. This is the von Neumann time step 
\begin{equation}
\Delta t_\mathrm{VC} = \frac{C_\mathrm{CFL}}{\pi^2 C_\mathrm{VC}^2 |\nabla\cdot \sV|},
\end{equation}
where $C_\mathrm{VC}$ is a constant, which we usually take to be $2$. 
The arbitrary form of the constant term $\pi^2 C_\mathrm{VC}^2$ comes from an analogy 
with the form of the von Neumann time step constraint by considering
the von Neumann term
\begin{equation}
C_\mathrm{VC}^2\lambda^2(-\nabla\cdot \sV)_{+}(-\nabla\cdot \sV)
\end{equation}
where $()_+$ denotes that the expression is zero if the term contained is negative,
as a diffusion operator and following the time step constraint from \citet[][Eq. 8]{2009ApJS..182..468M}.
We also introduce a similar constraint based on the shear of the flow
to allow for needed regularization, 
although the constraint on this vorticity time step much looser than in compression and expansion:
\begin{equation}
\Delta t_\mathrm{VR} = \frac{C_\mathrm{CFL}}{10 \pi^2 C_\mathrm{VC}^2 |\nabla \times \sV|}.
\end{equation}
The factor $10 \pi^2 C_\mathrm{VC}^2$ is an ad-hoc scaling
that in practice has been found to be sufficient.

The time step limit assigned for a particle is 
\begin{equation}
\Delta t = \min (\Delta t_{\mathrm{CFL}}, \Delta t_{\zeta},\Delta t_{\mathrm{VC}},
\Delta t_{\mathrm{VR}}).
\end{equation}
Each particle has an individually assigned time step. 
To tailor the algorithm to parallel implementation, a block time step scheme can be
adopted.
For such a time step scheme, the time steps actually used are rounded
down to the nearest block time step interval. 

It is also necessary to ensure some degree of spatial coherence
to the time steps, so that disturbances propagate from short-time step
particles to long-time step particles smoothly.
At the end of the time update procedure for a particle, after 
the assignment of the new time step for a particle,
if any of the neighbor particles
has an end time greater than the target particle's new end time, a time step limit
propagation procedure is triggered for the target particle.
The target particle's end time propagates
to its neighbors, and if any neighbor's end time is
further from 
the current time than twice the 
interval to the target particle's end time, then the neighbor's end time is set to this limit.

\section{Artificial Diffusion} \label{sec_artificialdiffusion}

Three types of artificial diffusion terms are used to stabilize the
solutions to the equations modeled in Phurbas.
These are terms acting as bulk viscosity, mass and internal energy diffusion terms, and a term acting to diffuse magnetic monopoles.
We find that these terms are sufficient to damp small scale 
fluctuations that would otherwise make the scheme unstable.

\subsection{Bulk Viscosities} \label{sec_bulkviscosities}

We use two bulk viscosity fields.
The first, $\zeta_l$, quenches small scale compressive motion in all areas of the flow, and 
evolves according to 
Equation~(\ref{eqzetal}).
The second, $\zeta_s$, is a shock and discontinuity viscosity
that evolves according to 
Equation~(\ref{eqzetas}).
Equations~(\ref{eqzetal}) and~(\ref{eqzetas}) each consist of three terms,
a diffusion term, a source term, and a decay term.
This configuration ensures that the bulk viscosity fields vary smoothly in time and space.

The diffusion operator on the bulk viscosity fields $\kappa_\zeta =
0.15 \lambda c_\mathrm{max}$ is chosen to  
place a time step limit less stringent than the Courant
limit from the hyperbolic part of the MHD equations.
The bulk viscosity fields have source and decay terms and these terms have associated timescales.
For the $\zeta_l$ field the source and decay term timescales are the same, $\tau_l= \lambda/c_\mathrm{max}$.
The $\zeta_s$ field is designed to fall off more slowly than it rises, which is 
particularly advantageous calming post-shock oscillations, so $\tau_{s-}= 20\lambda/c_\mathrm{max}$ and
$\tau_{s+}= \lambda/c_\mathrm{max}$.
The use of a diffusion equation with source and decay terms to derive the artificial viscosity field here is analogous
to the design in a discontinuous Galerkin method by \cite{2010JCoPh.229.1810B} 
and the slow decay of the shock viscosity achieves a similar effect to
the bulk viscosity prescription used by \cite{1997JCoPh.136...41M}. 

The $\zeta_s$ source term $S_s$ is given by:
\begin{align}
S_s = &\max\left( C_{VN}^2(-\partial_i V_i)_+ \lambda^2, \right. \nonumber \\
&C_e |\partial_i\partial_i (\sigma/\rho)|\lambda^2, \nonumber \\
&C_\rho \rho^{-1}|\partial_i\partial_i (\rho)|\lambda^3 \sqrt{c_s^2 +
  v_a^2}, \nonumber \\
&\left. C_P P^{-1}|\partial_i\partial_i (P)| \lambda^3 \sqrt{c_s^2 +
    v_a^2}\right). 
\end{align}
The first term has the form of the conventional von Neumann artificial viscosity, with 
$C_{VN}=2$.
The second term responds to changes in the specific internal energy, in a manner similar to
the dissipation introduced by \cite{2008JCoPh.22710040P}. 
This form is a trigger on the size of the second derivative of the specific internal energy, and
$C_{e}=0.1$.
The third term is constructed analogously to the second, but using the Laplacian of density and 
$C_{\rho}=1.0$.
The final term is again constructed analogously to the second, but using the Laplacian of pressure and 
$C_{P}=3.0$.
The constants $C_{VN}$, $C_e$, and $C_P$ can be tuned for a particular problem, with smaller values being preferable,
but the values given here have proven to be sufficient for most problems.

The mass and internal energy diffusion terms are coupled to the 
$\zeta_s$ field, with a strength set by the constants
$H_\rho=H_\sigma=5\times10^{-4}$. 
For stability, it would be preferable to have a small scale mass and internal energy diffusion
(such as a hyperdiffusion) active everywhere in the flow, but we have not found a formulation
 of such a term that is sufficiently accurate to yield reasonable mass conservation results.

\subsection{Magnetic Divergence Diffusion} \label{sec_magneticdivergence}

As Phurbas solves the equations of MHD, the issue of magnetic monopole errors
must be treated.  The primary problem caused by monopole errors in schemes of
this type is numerical instability. 
The interpolating moving least squares derivative estimates may return derivatives of the
magnetic field that do not satisfy $\nabla \cdot \B = 0$.
Over several time update cycles, these estimates may 
lead to the local creation of a net magnetic monopole character to the field.  
In practice we have found that a diffusive (or parabolic) correction is sufficient 
to prevent the growth of this monopole character of the magnetic field (see tests in Paper II).

For each particle, a $\divb$ field is defined. The value is simply
reset each time step to the value of $\divb$ derived from the fit to
the magnetic field.  The derivatives of the $\divb$ field derived from
the fits are then used to diffuse $\divb$, generally resulting in a
reduction of its value. These fitted values and derivatives of $\divb$
are less noisy than values and derivatives of $\divb$ derived directly
from fits to the magnetic field.

The diffusion term for particle $j$ is
\begin{equation}
\xi_j = \eta_{max} \nabla(\nabla \cdot \B_j), \label{eqdivbdiffusion}
\end{equation}
where $\eta_{max}$ is the maximum diffusion coefficient possible under 
the stability criterion
\begin{equation}
\Delta t < \frac{\lambda^2}{\pi^2 \eta}
\end{equation}
from \citet[][Eq. 8]{2009ApJS..182..468M}.
The term given by Equation~(\ref{eqdivbdiffusion}) is added to the
right hand side of the induction equation (Equation~\ref{eqinduction}).
in the first time derivatives used in the second-order predictor-corrector 
scheme for the evolution of the magnetic field. 
The effect of this is that the $\divb$ diffusion operator is 
integrated with a first-order predictor-corrector scheme.
The $\divb$ diffusion $\eta_{max}$ is computed each time the fields are fit,
which occurs at times that are the end of one time step and the beginning of the next.
The time step used to define $\eta_{max}$ is the time step that has
its end at the instant $\eta_{max}$ is calculated, i.e. the  previous time step.
Thus, the $\eta_{max}$ used in the predictor stage of the time
integration of a particular step is 
different from the $\eta_{max}$ later used in the corrector stage of
the same time step. 
This diffusion is not
conservative, but the Phurbas discretization only preserves the conservation in
the MHD equations to truncation error levels, and the $\divb$ operated on by
this diffusion is, by definition, only created below truncation error levels.  We
find that since the canonical form of the Lagrangian MHD equations
     that we use 
treats $\divb$ as a passively advected scalar, the presence of small
amounts of $\divb$ does not destabilize the solution. 

\section{Summary and Discussion} \label{sec_discussion}

\subsection{Summary of the Algorithm} \label{sec_summary}

We now summarize the conceptual steps of the algorithm.
The operations described here are actually often broken into
multiple phases to enable efficient parallelization.
Future implementers of the algorithm should consider the specific
needs of each operation when designing data structures and
communication patterns. 

\begin{itemize}

\item Build the neighbor-finding data structure, such as a particle tree.
\item Balance particles among processors.
\item Identify target particles for evolution.
\item Use the tree to assemble all neighbors within a radius $r_f$ of the
target particles.
\item Check for 
   voids.
Complete particle addition for qualifying voids (\S~\ref{sieve}).
\item Check if there are mutual nearest neighbor pairs that are too close.
Delete one 
 (\S~\ref{sec_clumpbusting}).
\item Evaluate the polynomial fit weights (\S~\ref{sec_weights}).
\item Compute the local polynomial fit 
  to derive values and spatial derivatives at the location of each
  particle
(\S~\ref{sec_matrix}).
\item Use the polynomial coefficients to evaluate the MHD equations 
         for the Lagrangian time derivatives including diffusivity terms (\S~\ref{sec_algorithm}).
\item Use the polynomial coefficients to evaluate the governing equations of the bulk viscosity fields
         $\zeta_l$ and $\zeta_s$ 
         for the Lagrangian time derivatives (\S~\ref{sec_algorithm}).
\item Use the time derivatives to correct the previous time step
(\S~\ref{sec_timeupdate}).
\item Evaluate the resolution scale $\lambda$ for each particle
  position (\S~\ref{sec_algorithm}).
\item Evaluate the size of the next time step (\S~\ref{sec_timestep}).
\item Use the time derivatives to predict forward in time to the
next time step (\S~\ref{sec_timeupdate}).
\item Restrict local time step variations (\S~\ref{sec_timestep}).
\end{itemize}

\subsection{Effective Resolution} \label{effectiveresolution}

To understand the effect of varying the effective resolution parameter
$\lambda$ on the numerical resolution, consider a one-dimensional
uniform grid with a grid spacing of unity, initialized with a field
variable having values given by the Fourier mode
$\sin(\pi k x).$ The maximum wave number $k$ of this mode that can be
expressed on this grid is $k=1$, the Nyquist wave number. 
In order to
be able to calculate realistic derivatives with finite differences,
$k$ must be less than unity, and the precision increases as $k$
decreases. \citet{2008ApJ...677..520M} and \citet{2009ApJS..182..468M}
evaluate the effective precision of finite difference schemes with
varying stencil sizes.
They find that for a stencil radius of $\{1,2,3,4\},$ finite differences can be
calculated with a relative precision of $\sim 1$ percent up to a wave
number of $k \sim \{1/8,1/4,2/5,1/2\}.$ Given that derivatives are more
easily calculated on a grid than for irregular particles, we take this
as an upper limit for what we can expect from particles.
Since a 3D, third-order, polynomial corresponds to a 5-point (or
stencil radius 2) 1D finite-difference scheme,
we expect $k \sim 1/4$ to be the limit of resolution, corresponding
to a wavelength of $\sim 8 \lambda$.

\section{Conclusion}
We have described Phurbas, an adaptive, Lagrangian, meshless algorithm
for MHD.  The algorithm is described for the specific case of the MHD
equations, but can be easily generalized to other hyperbolic systems,
as the fitting, time integration, and stabilization procedures do not
rely on particular properties of the MHD equations.  The central
principle of the algorithm is that the solution and its spatial
derivatives are derived from a high-order, interpolating, polynomial fit to a set of
particles that are merely Lagrangian sample points in the flow, not
mass elements as in SPH or finite volume methods.  This allows for
significant flexibility in the design of the algorithm, and the
implementation of additional physical processes.  
Particle addition and deletion is required to prevent
the growth of voids or clumps. This naturally allows the numerical resolution
to be fully adaptive based on user specified criteria.  The
Lagrangian nature of the code means that particles can be evolved
with time steps dependent only on the nature of the local flow, and
that numerical diffusion is Galilean invariant.
The version
described here is just one subset of the many available options.
Paper II describes a parallel implementation and tests of Phurbas 
that demonstrate accuracy comparable to that of third order grid
codes on subsonic and supersonic problems.

A few theoretically desirable improvements to the scheme can already be identified. 
Though the $\zeta_l$ field is sufficient to modify the equations to give reasonable results in 
tests, it would be preferable to only use stabilizing viscosities
that scale as $\lambda^2$ or a higher power (hyperviscosities) to give
faster convergence of the modeled equations 
to the limit of ideal MHD.  
Possible avenues though which such an effect could be achieved
are use of a different interpolation scheme than moving least squares,
 and/or a time or spatially dependent version of $\zeta_l$ so that it couples only to the
shortest wavelength or
shortest timescale motions.
\cite{KuhnertThesis} showed methods for introducing upwinding into a scheme similar to Phurbas, 
a modification that may reduce the need for $\zeta_l$.
Additionally, in defining $\zeta_s$ we have used a simple formulation where the effects decay on the same 
timescale and reflect the effects of discontinuities (shocks and otherwise) through the same parameter.
In SPH, \cite{2012JCoPh.231..759P} and \cite{2009NewAR..53...78R} 
have found that separate parameters for each discontinuity are
useful. Analogously in the framework here, $\zeta_s$ could be broken into three fields, 
though this would come at some cost.

The second order Hermite predictor-corrector scheme is particularly useful as 
it only requires computations of spatial derivatives at the beginning of each time step, 
and the predictor half of the integration can be done without knowing the end-time of the time step.
It however has the drawback that first and second time derivatives of the field variables
must be obtained. 
In general, these analytic expressions could be very complicated.
A time integration scheme, such as a Runge-Kutta method,
that uses only first time derivatives may be preferable in this sense.

\acknowledgments The anonymous referee made a contribution equivalent
to authorship, by pointing out the need for the approximating functions to be
interpolating to maintain numerical stability, and by providing a
first version of the stability analysis given in Appendix A. The
authors also acknowledge useful discussions with A. Brandenburg,
V.~Springel and S.C.O.~Glover.
M.-M.M.L.~and
C.P.M.~acknowledge hospitality from the Max-Planck-Institut f\"ur
Astronomie, and M.-M.M.L~additionally acknowledges hospitality of the
Institut f\"ur Theoretische Astrophysik der Uni.\ Heidelberg.  This
work has been supported by National Science Foundation grants
AST-0612724 and AST-0835734.  Computations were performed under allocation
TG-MCA99S024 from the Extreme Science and Engineering Discovery
Environment (XSEDE), which is supported by National Science Foundation
grant number OCI-1053575.

\clearpage

\appendix

\section{Stability of a Model Scheme} \label{sec_stability}

\begin{figure}
\plotone{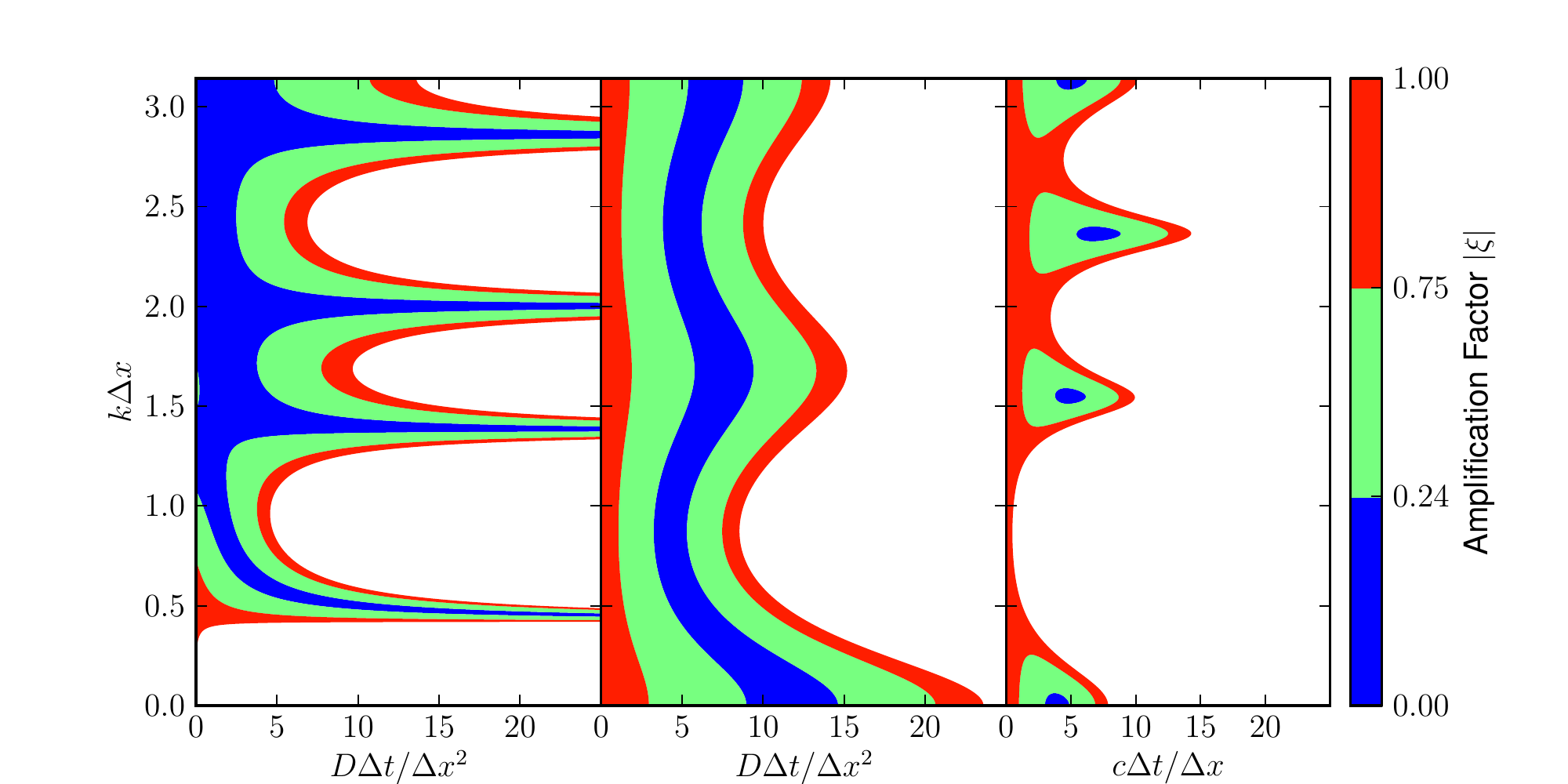}
\caption{ Von Neumann stability analysis of model schemes, showing the
  amplification factor $|\xi|$ as a function of perturbation
  wavelength $k$ and time step $\Delta t$, both appropriately
  normalized to the grid.  The unstable region with values of $|\xi| >
  1$ is shown in white. {\em Left:} the \citet{2003ApJ...595..564M}
  type scheme for the diffusion equation.  Note the instability of low
  wavenumber perturbations $k\Delta x \lesssim 0.4$.  {\em Middle:}
  Least squares interpolant scheme for the diffusion equation.  {\em
    Right:} Least squares interpolant scheme for the
  advection-diffusion equation.  Both the latter schemes show a region
  of stability at small enough time step for all wavenumber
  perturbations.  }
\label{figmodelstab}
\end{figure}

To determine the stability requirements for our meshless method, we
here present a stability analysis for a simple model that captures the
essential points of our algorithm.
We analyze a numerical scheme for approximating the solution of the diffusion equation
\begin{align}
\frac{\partial u}{\partial t} = D\frac{\partial^2 u}{\partial x^2} \label{eqdiffusionproblem}.
\end{align}
We discretize $u(x)$ on a grid with spacing $\Delta x$.
To find derivatives of $u$ we use a moving least squares approximation. 
Shifting the origin to grid point $x_0$ the polynomial approximation of
$u(x)$ is given by
\begin{align}
U(x) &= \sum_{p=0}^{P} a_px^p \label{eqpolyform}.
\end{align}
The coefficients $a_p$ are determined by minimizing the sum of the square of errors $E_2$
 at $2N+1$ neighboring grid points, given by
\begin{align}
E_2 &= \sum_{j=-N}^N (u_j-U(x_j))^2.
\end{align}
Combining these two expressions, and using the definition of the grid point 
positions $x_j = j \Delta x$ gives
\begin{align}
E_2 &= \sum_{j=-N}^N (u_j- \sum_{p=0}^{P} a_p j^p \Delta x^p  )^2.
\end{align}
We can find the  the minimum of $E_2$ by
setting $\partial E_2/\partial a_q=0$ yielding
\begin{align}
\sum_{j=-N}^N j^q \Delta x^q u_j = \sum_{j=-N}^N \sum_{p=0}^P a_p j^{p+q} \Delta x^{p+q},
\end{align}
which is a system of equations for $q=0..P$ that can be solved for the polynomial coefficients $a_p$.
The second derivative of the polynomial Equation~(\ref{eqpolyform})  at
$x=0$ is $2a_2$.
So, we can write a scheme to update the solution to Equation~(\ref{eqdiffusionproblem}) as
\begin{align}
u^{n+1}_0 = a_0 +2D a_2\Delta t,
\end{align}
where $u_0 = u(x_0)$. In each step we replace the value $u_0^{n}$ with
the fit $a_0$ and use the second derivative of the moving least squares fit to
construct a forward-Euler type time update. This is a
 \citet{2003ApJ...595..564M} type scheme for solving the diffusion equation.
As an example, for third order polynomials ($p=3$) and $N=4$ this scheme is specifically
\begin{align}
u^{n+1}_0 &= u_0\left(\frac{59}{231} - \frac{10}{231}\frac{D \Delta t}{\Delta x^2}\right)
+\frac{59}{231} \sum_{j=1}^{4}\left(u_j+u_{-j}\right) -\frac{5}{231}\sum_{j=1}^{4} j^2\left(u_j+u_{-j}\right)\\
&+\frac{D \Delta t}{\Delta x^2}\left[\frac{-1}{231} \sum_{j=1}^4 \left(u_j+u_{-j}\right) +\frac{1}{154}\sum_{j=1}^4 j^2(u_j+u_{-j})\right].
\end{align}
To perform a von Neumann stability analysis, we substitute $u_\ell^n = \xi^n e^{ik\ell\Delta x}$, and solve for $\xi$ 
\begin{align}
\xi &= \frac{1}{231} \left(59 - 10\frac{D \Delta t}{\Delta x^2} + \sum_{j=1}^4 (59-5 j^2)\cos(jk\Delta x) \right)\nonumber\\
 & +\frac{D \Delta t}{\Delta x^2} \sum_{j=1}^4 \left( \frac{-1}{231} +\frac{j^2}{154}\right)\cos(jk\Delta x).
\end{align}
We can evaluate this expression numerically, plotting the amplification factor $|\xi|$ 
for each wavenumber $k\Delta x$ and time step $\frac{D \Delta t}{\Delta x^2}$.
This is shown in Figure~(\ref{figmodelstab}). 
The fundamental trouble with \citet{2003ApJ...595..564M} type schemes using a moving least squares fit 
is the region at low wavenumbers and small time steps where the magnitude of the amplification factor is
greater than unity. In this region the scheme is unstable.

If instead we use a moving least squares interpolant instead of just a
fit, the system of equations for the coefficients is given by 
\begin{align}
\sum_{j=-N}^N j^q \Delta x^q\left(u_j-u_0\right) &= \sum_{j=-N}^N \sum_{p=1}^P a_p j^{p+q} \Delta x^{p+q}
\end{align}
for $q=1,2,3$.
Here we have eliminated the coefficient $a_0$ by forcing the moving least squares approximation to interpolate $u_0$.
Then we can write a forward-Euler type scheme using the second derivative of this interpolant as
\begin{align}
u^{n+1}_0 =u^{n}_0 +2D a_2\Delta t.
\end{align}
For $p=3$ and $N=4$ this gives the scheme
\begin{align}
u^{n+1}_0 &= u_0 + \frac{D \Delta t}{354 \Delta x^2} \sum_{j=0}^4 j^2\left(u_j+u_{-j} -2 u_0 \right)\\
&= u_0\left(1- \frac{60 D \Delta t}{354 \Delta x^2} \right)+ \frac{D \Delta t}{354 \Delta x^2} \sum_{j=1}^4 j^2\left(u_j+u_{-j} \right).
\end{align}
Again, performing a von Neumann stability analysis, we substitute $u_\ell^n = \xi^n e^{ik\ell\Delta x}$ and solve for $\xi$
\begin{align}
\xi &= \left(1-\frac{60D\Delta t}{354\Delta x^2}\right) +\frac{D\Delta t}{354 \Delta x^2} \sum_{j=1}^{4} j^2 2\cos(jk\Delta x).
\end{align}
The magnitude of $\xi$ in this case is shown in the center panel of Figure~(\ref{figmodelstab}) for the case $D=c\Delta x$. 
In stark contrast to the  \citet{2003ApJ...595..564M} type scheme constructed with a non-interpolating, moving least squares fit,
a significant region of stability ($|\xi|<1$) exists at small timestep
for all wavenumbers $k\Delta x$. 

For the advection equation
\begin{align}
\frac{\partial u}{\partial t} = c\frac{\partial u}{\partial x} + c\Delta x \frac{\partial^2 u}{\partial x^2} \label{eqadvectdiffusionproblem}.
\end{align}
we can write the scheme
\begin{align}
u^{n+1}_0 =u^{n}_0 +c\Delta t a_1 +2 c a_2 \Delta x \Delta t.
\end{align}
Here we have set the diffusion parameter to be scaled by the grid resolution $\Delta x$.
With the interpolating moving least squares coefficients $a_1$ and $a_2$ as in the diffusion problem  above we have
\begin{align}
u^{n+1}_0  &= 
u_0\left(1- \frac{60 c\Delta x \Delta t}{354 \Delta x^2} \right)
+\frac{c\Delta t}{7128 \Delta x}\left(59\sum_{j=1}^{4}j^2(u_j-u_{-j}) - 815\sum_{j=1}^{4}j(u_j-u_{-j})\right) \nonumber\\
&+ \frac{c\Delta x \Delta t}{354 \Delta x^2} \sum_{j=1}^4 j^2\left(u_j+u_{-j} \right)
\end{align}
A von Neumann stability analysis yields
\begin{align}
\xi &= \left(1-\frac{60c\Delta t}{354\Delta x}\right) 
+\frac{i c\Delta t}{7128 \Delta x}\left(59\sum_{j=1}^{4}2 j^2\sin(jk\Delta x) - 815\sum_{j=1}^{4}2 j\sin(jk\Delta x) \right) \nonumber\\
& +\frac{c\Delta t}{354 \Delta x} \sum_{j=1}^{4} j^2 2\cos(jk\Delta x).
\end{align}
Note that the second term is imaginary, arising from the advection
operator, so if the contributions from the diffusion operator, the
part of the first term, and the final term, were dropped, 
the scheme for the pure advection problem would be unconditionally unstable.
We plot the magnitude of the amplification factor $|\xi|$ in Figure~(\ref{figmodelstab}).
The addition of the diffusion operator has stabilized the advection
problem for sufficiently small time steps $c \Delta t / \Delta x$.

\section{Time Integration} \label{app_timeintegration}

Time integration proceeds using a system of Hermite
predictor-corrector formulas for field variable and position updates.
This scheme is a lower order version of that presented in
\citet{2008NewA...13..498N}.  As it is has not previously been
described in the literature, we present here a brief derivation of the
integration method.

We predict field values $q_{p,i+1}$ at time $t+\Delta t$
using a Taylor series expansion around their values at time $t$ incorporating
their time derivatives calculated after the prediction phase of the
previous time step, giving
\begin{align}
q_{p,i+1} = q_{c,i} + D_{t}q_{p,i} \Delta t + \frac{1}{2}D_{tt}q_{p,i} \Delta t^2 
\end{align}
where $q_{c,i}$ is the corrected field value from the previous time
step at time $t$.  After calculating the values of $q_{p,i+1}$, we use
them to calculate $D_tq_{p,i+1}$ and $D_{tt}q_{p,i+1}$, as given by
Equations~(\ref{eqmomentumstab})--(\ref{eqzetas}),
and those in Appendix~\ref{app_secondderiv}.  These derivatives will be used in
the corrector stage of the current time step and in the predictor
stage of the next time step.
The stabilizing diffusion terms, linked to the $\zeta_s$ and $\zeta_l$ fields
are proportional to the resolution parameter $\lambda$. 
As asymptotically, the time step
 varies as $\lambda$ itself due to the CFL limit, we only 
integrate these terms to first order in time, so the 
combined space-time error is second order in $\lambda$.

By using higher time derivatives, a Hermite scheme of time integration
depending only on field variable values from the previous time step
can be constructed.  This saves storage, avoids complex start up
procedures, and simplifies the use of individual particle time steps in
comparison to predictor-corrector schemes based on Newton
interpolation (such as Aarseth and Adams-Bashforth-Moulton schemes) that
require storage of field values from earlier time steps.  The Hermite
corrector stage is constructed as
\begin{align}
q_{c,i+1} = q_{c,i} +\int_0^{\Delta t} f_c(\tau) d\tau. \label{eqcorrform}
\end{align}
We choose the function $f_c(\tau)$ to be a Hermite interpolation, that
is, a polynomial that interpolates $D_t q$ and $D_{tt} q$ at each end
of the time step.  We further choose to simplify the formalism by
designing the polynomial to be time symmetric about $t+\Delta t/2$, so that
\begin{align}
f_c(\tau) = f_0 +f_1\left(\tau-\frac{\Delta t}{2}\right) + f_2\left(\tau-\frac{\Delta t}{2}\right)^2 + f_3\left(\tau-\frac{\Delta t}{2}\right)^3.
\end{align}
We determine the coefficients $f_0,\ f_1,\ f_2,\ f_3$ by using four
constraints: at $\tau = 0$, $f_c(\tau)$ must have a value of $D_t q_{p,i}$,
and a time derivative $D_{tt} q_{p,i}$; while at $\tau = \Delta t$ the
value and the derivative must be $D_t q_{p,i+1}$ and $D_{tt}
q_{p,i+1}$, respectively. However, evaluating the integral in Equation~(\ref{eqcorrform}), 
the time symmetry we chose yields the simple result that the
$f_1$ and $f_3$ terms integrate to zero regardless of the values of
their coefficients.  Performing the integral in
Equation~(\ref{eqcorrform}) yields a correction stage
\begin{align}
q_{c,i+1} = q_{c,i} + \frac{1}{2}(D_t q_{p,i} + D_t q_{p,i+1} ) \Delta
t +\frac{1}{12}(D_{tt} q_{p,i} - D_{tt} q_{p,i+1}) \Delta t^2.
\end{align}

Velocity $V$ is treated as an independent set of field variables, so
for particle positions $x$ there are three time derivatives of
information available, as well as corrected values of velocity from
the beginning and end of the current time step.
Therefore, for the predictor stage for the position we use a third
order Taylor series incorporating the best information available at
time $t$,
\begin{align}
x_{p,i+1} = x_{c,i} + V_{c,i} \Delta t + \frac{1}{2}D_{t}V_{p,i} \Delta t^2 + \frac{1}{6}D_{tt}V_{p,i} \Delta t^3.
\end{align}
Similarly to the field variable integration we choose a
time-symmetric, Hermite interpolating function, so that the corrector stage is
\begin{align}
x_{c,i+1} = x_{c,i} +\int_0^{\Delta t} g_c(\tau) d\tau. \label{eqxcorrform}
\end{align}
The function $g_c(\tau)$ is again a polynomial centered on $t+\Delta
t/2$, that now interpolates through $V_{c,i}$, $D_t V_{p,i}$, and
$D_{tt} V_{p,i}$ at $\tau=0$, and through $V_{c,i+1}$, $D_t V_{p,i+1}$,
and $D_{tt} V_{p,i+1}$ at $\tau=\Delta t$. (The availability of
$V_{c,i+1}$ occurs because we have, at this point, already updated the
field variables.) Applying these constraints allows us to evaluate the
coefficients in the interpolating polynomial
\begin{align}
g_c(\tau) = g_0 + g_1(\tau-\frac{\Delta t}{2}) + g_2(\tau-\frac{\Delta t}{2})^2 + g_3(\tau-\frac{\Delta t}{2})^3 + g_4(\tau-\frac{\Delta t}{2})^4 + g_5(\tau-\frac{\Delta t}{2})^5.
\end{align}
Evaluating the integral in Equation~(\ref{eqxcorrform}), the $g_1$, $g_3$, and $g_5$ terms are 
zero due to the choice of time symmetry, and so the correction stage
for particle positions is
\begin{align}
x_{c,i+1} = x_{c,i} + \frac{1}{2}(V_{c,i} + V_{c,i+1} ) \Delta t +\frac{1}{10}(D_{t} V_{p,i} - D_{t} V_{p,i+1}) \Delta t^2 +\frac{1}{120}(D_{tt} V_{p,i} + D_{tt} V_{p,i+1}) \Delta t^3.
\end{align}

It can be useful to split the integration in to a background flow and
perturbations, for example in computing the dynamics of a steady-state cylindrical flow.
In this case we define the perturbation velocity field as $\sV' = \sV - \Omega r\hat{\mathsf{\phi}}$
where $r$ is the two-dimensional radius from the center of the cylinder, and $\Omega(r)$ is
the angular velocity of the background flow.
We denote the components of the circular radius to the point $(x_{c,i},y_{c,i})$ as $r_{c,i,x},r_{c,i,y}$,
and the angular velocity of the background flow at this radius is $\Omega_p$.
Then, the predictor step with the background flow separated from the perturbation velocity $\sV'$ is
\begin{align}
x_{p,i+1} = r_{c,i,x}\cos(\Omega_p \Delta t) - r_{c,i,y}\sin(\Omega_p \Delta t) + V'_{c,i,x} \Delta t + \frac{1}{2}D_{t}V'_{p,i,x} \Delta t^2 + \frac{1}{6}D_{tt}V'_{p,i,x} \Delta t^3\\
y_{p,i+1} = r_{c,i,y}\cos(\Omega_p \Delta t) + r_{c,i,x}\sin(\Omega_p \Delta t) + V'_{c,i,y} \Delta t + \frac{1}{2}D_{t}V'_{p,i,y} \Delta t^2 + \frac{1}{6}D_{tt}V'_{p,i,y} \Delta t^3 .
\end{align}
We then add the background flow state back into the field variables used in Phurbas
to calculate time derivatives for all fields in the usual inertial reference frame
with $\sV$ not $\sV'$.
Then, to transform the time derivatives to time derivatives of the perturbation velocity we use
\begin{align}
D_t V'_{p,i+1,x} &= D_t V_{p,i+1,x} + \Omega^2_p r_{p,i+1,x}\\
D_t V'_{p,i+1,y} &= D_t V_{p,i+1,y} + \Omega^2_p r_{p,i+1,y}\\
D_{tt} V'_{p,i+1,x} &= D_{tt} V_{p,i+1,x} - \Omega^3_p r_{p,i+1,y}\\
D_{tt} V'_{p,i+1,x} &= D_{tt} V_{p,i+1,x} + \Omega^3_p r_{p,i+1,x}. 
\end{align}
The perturbation velocity $\sV$ can be integrated directly using these perturbation time derivatives.
Using the normal corrector for the field variables and the
perturbation velocity, the corrector step for position is then
\begin{align}
x_{c,i+1} = r_{c,i,x}\cos(\Omega_p \Delta t) - r_{c,i,y}\sin(\Omega_p \Delta t) + \frac{1}{2}(V'_{c,i,x} + V'_{c,i+1,x} ) \Delta t +\frac{1}{10}(D_{t} V'_{p,i,x} - D_{t} V'_{p,i+1,x}) \Delta t^2 \nonumber\\
+\frac{1}{120}(D_{tt} V'_{p,i,x} + D_{tt} V'_{p,i+1,x}) \Delta t^3 \\
y_{c,i+1} = r_{c,i,y}\cos(\Omega_p \Delta t) + r_{c,i,x}\sin(\Omega_p \Delta t) + \frac{1}{2}(V'_{c,i,y} + V'_{c,i+1,y} ) \Delta t +\frac{1}{10}(D_{t} V'_{p,i,y} - D_{t} V'_{p,i+1,y}) \Delta t^2 \nonumber\\
+\frac{1}{120}(D_{tt} V'_{p,i,y} + D_{tt} V'_{p,i+1,y}) \Delta t^3 .
\end{align}
To prepare for the next step, it is necessary to shift the perturbation velocity and time derivatives into
the correct frame that will be used for the next step.
The new frame at the corrected position $(x_{c,i+1},y_{c,i+1})$ has radius components $(r_{c,i+1,x},r_{c,i+1,y})$
and the angular velocity of the background flow at this radius is $\Omega_c$.
The transformations made to these quantities are:
\begin{align}
V'_{c,i+1,x} &\rightarrow V'_{c,i+1,x} - r_{p,i+1,y} \Omega_p + r_{c,i+1,y} \Omega_c\\
V'_{c,i+1,y} &\rightarrow V'_{c,i+1,y} + r_{p,i+1,x} \Omega_p - r_{c,i+1,x} \Omega_c\\
D_t V'_{c,i+1,x} &\rightarrow D_t V'_{c,i+1,x} - r_{p,i+1,x} \Omega_p^2 + r_{c,i+1,x} \Omega_c^2\\
D_t V'_{c,i+1,y} &\rightarrow D_t V'_{c,i+1,y} - r_{p,i+1,y} \Omega_p^2 + r_{c,i+1,y} \Omega_c^2\\
D_{tt} V'_{c,i+1,x} &\rightarrow D_{tt} V'_{c,i+1,x} + r_{p,i+1,y} \Omega_p^3 - r_{c,i+1,y} \Omega_c^3\\
D_{tt} V'_{c,i+1,y} &\rightarrow D_{tt} V'_{c,i+1,y} - r_{p,i+1,x} \Omega_p^3 + r_{c,i+1,x} \Omega_c^3
\end{align}
This shifts $\sV'$ into the accelerating reference frame used for the following predictor step.

\section{Second Time Derivatives of MHD Equations} \label{app_secondderiv}

For the second order predictor-corrector scheme \S~\ref{sec_timeupdate} 
we need both the first and second Lagrangian time derivatives of the
MHD equations~(\ref{eqmomentum})--(\ref{eqmass}).
This appendix gives formulas for the required second time derivatives.

The form for a Lagrangian time derivative $D_t$ in terms of partial
derivatives $\partial$ is: 
\begin{align}
D_t \partial q &=\partial_t \partial q + V\cdot\nabla q\\
 & = \partial_t \partial q + \partial[V\cdot\nabla q] - \partial V_i \partial_i q \\
 & = \partial_t\partial_j q + \partial_j[V_i\partial_i q] - (\partial_j V_i)(\partial_i q) \\
 & = \partial D_t q - (\partial V_i)(\partial_i q)
\end{align}
Applying this to the MHD equations~(\ref{eqmomentum})--(\ref{eqmass})
gives the second time derivatives needed for the 
second order predictor corrector scheme.
We start with the velocity equation (Equation~\ref{eqmomentum}). 
Taking its Lagrangian time derivative, the first term on the right hand side becomes
\begin{align}
D_{t}(-\rho^{-1}\partial_j P)  =  -\rho^{-1} (\partial_jD_t P - \partial_j V_a\partial_a P) + \rho^{-2}(D_t\rho)\partial_j P .
\end{align}
Inserting this, the second derivative of velocity is
\begin{align}
D_{tt} V_j = -\rho^{-1}\left(\partial_j D_tP - \partial_j V_i\partial_iP\right) + \rho^{-2}(\partial_jP) D_t P + D_{t}(\rho^{-1} (\varepsilon_{jab}\varepsilon_{acd}(\partial_c B_d) B_b)).
\end{align}
This equation can be further reduced.  The two pressure-dependent terms depend on the equation
of state. For a gamma-law equation of state,
$P=(\gamma-1)\sigma$,
\begin{align}
D_tP = (\gamma-1)D_t\sigma = (\gamma-1)(-(\sigma+P)\partial_i V_i) ,
\end{align}
and so
\begin{align}
\partial_jD_t P =
(\gamma-1)\left(-(\partial_j\sigma+\partial_jP)\partial_a V_a
  -(\sigma+P)\partial_{aj} V_a\right). 
\end{align}
For an isothermal equation of state $P=c_s^2\rho$,
\begin{align}
D_tP = c_s^2 D_t\rho = c_s^2 (-\rho\partial_i V_i), 
\end{align}
and so
\begin{align}
\partial_jD_t P = c_s^2\left(-\partial_j\rho\partial_a V_a - \rho\partial_{aj} V_a\right).
\end{align}
The magnetic term reduces to
\begin{align}
D_{t}&(\rho^{-1} (\varepsilon_{jab}\varepsilon_{acd}(\partial_c B_d)
B_b))  = (\varepsilon_{jab}\varepsilon_{acd}(\partial_c B_d)
B_b)(-\rho^{-2}D_t\rho) \nonumber \\ 
   & +
   \rho^{-1}\left(\varepsilon_{jab}\varepsilon_{acd}\left(\left((\partial_c
         B_e)\partial_e V_d +  B_e\partial_{ce} V_d -(\partial_c
         B_d)\partial_e V_e - B_d \partial_{ce}V_e -(\partial_c
         V_e)\partial_e B_d\right) B_b \right.\right.\nonumber\\ 
  & \left. \left. + (\partial_c  B_d)D_t B_b\right)\right),
\end{align}
while the Lagrangian time derivative of the gravitational force
\begin{align}
D_t(+G_j) = \partial_t G_j +  V_i\partial_i G_j.
\end{align}

Taking the Lagrangian time derivative of the induction equation
(Equation~\ref{eqinduction}) gives
\begin{align}
D_{tt}B_j = &(D_t B_i)\partial_i V_j + B_i(\partial_i D_t V_j -\partial_i V_k \partial_k V_j) \nonumber \\
& -(D_t B_j)\partial_i V_i -B_j(\partial_iD_t V_i - \partial_i
V_k\partial_k V_i) . 
\end{align}
The Lagrangian time derivative of the internal energy equation is
\begin{align}
D_{tt} \sigma & = -(\sigma + P)(\partial_iD_t V_i -\partial_i
V_j\partial_j V_i)-(D_t\sigma + D_t P)\partial_i V_i , 
\end{align}
where the required pressure-dependent expressions have already
appeared above. If we have a barotropic or isothermal equation of
state then this equation is not used, of course. 
Finally, the Lagrangian derivative of the continuity equation (Equation~\ref{eqmass}) is
\begin{align}
D_{tt}\rho = -(D_t\rho)(\partial_i V_i)-\rho(\partial_i D_t V_i -\partial_i V_j\partial_j V_i).
\end{align}
The second term on the right hand side can be expanded as
\begin{align}
\partial_iD_t V_i =& \varepsilon_{iab}\varepsilon_{acd}\left(-\rho^{-2}(\partial_c B_d)B_b \partial_i \rho + \rho^{-1}(\partial_{ic}B_d)B_b + \partial_c B_d\partial_i B_b\right)\nonumber \\
 &+\rho^{-2}\partial_iP\partial_i\rho - \rho^{-1}\partial_{ii}P + \partial_i G_i .
\end{align}

\end{document}